\documentclass[11pt]{article}

\usepackage[utf8]{inputenc} 
\usepackage[T1]{fontenc}    
\usepackage{url}            
\usepackage{amsfonts}       
\usepackage{amsmath}        
\usepackage{graphicx}       
\usepackage{geometry}       
\usepackage{subcaption}     
\usepackage{float}          

\graphicspath{ {./} }

\title{Generative AI Pipeline for Interactive Prompt-driven 2D-to-3D Vascular Reconstruction for Fontan Geometries from Contrast-Enhanced X-Ray Fluoroscopy Imaging: A Gemini 2.5 Flash ("Nano Banana")-Based Framework Enabling STL Generation and Virtual Hemodynamic Visualization}

\author{
 Prahlad G Menon, PhD, PMP \\
  University of Pittsburgh, Pittsburgh, PA \\
  The American Association for Thoracic Surgery (AATS)\\
  \texttt{prm44@pitt.edu} \\
}

\begin{document}
\maketitle

\begin{abstract}
Fontan palliation for univentricular congenital heart disease often progresses to hemodynamic failure characterized by elevated systemic venous pressures, reduced cardiac output, and complex flow patterns poorly characterized by conventional 2D imaging modalities. Current diagnostic assessment relies primarily on contrast-enhanced X-ray fluoroscopy angiography obtained during cardiac catheterization, which provides limited three-dimensional geometric information essential for computational fluid dynamics (CFD) analysis and evidence-based surgical planning. A multi-step AI pipeline was developed utilizing Google's Gemini 2.5 Flash multimodal large language model (2.5B parameters) for systematic, prompt-driven, iterative processing of contrast-enhanced fluoroscopic angiograms through transformer-based neural architecture with enhanced vision capabilities, culminating in 3D geometries suitable for geometric and computational fluid dynamics analysis.

The pipeline encompasses medical image preprocessing, vascular segmentation, contrast enhancement, artifact removal, 3D geometric optimization, and virtual hemodynamic flow visualization entirely within 2D projection renderings. A representative failing Fontan angiogram underwent systematic processing through prompt-driven refinement cycles with comprehensive quality control measures. Final optimized 2D projection views were processed through Tencent's Hunyuan3D-2mini diffusion transformer neural architecture (384M parameters) employing encoder-decoder networks with attention mechanisms, for stereolithography (STL) file generation.

The AI pipeline successfully generated geometrically optimized 2D projection views from single-view fluoroscopic angiograms after 16 processing steps using a custom web-browser-driven user interface. Initial iterations contained hallucinated vascular features requiring iterative prompt-driven regeneration cycles via the developed interface to achieve anatomically faithful representations. Final projection views demonstrated accurate preservation of complex Fontan geometric features with enhanced contrast suitable for 3D conversion. AI-generated virtual flow visualization identified stagnation zones in central total cavopulmonary connection and flow patterns in branch pulmonary arteries. Complete processing of this example required under 15 minutes total time, with individual edits appearing within seconds via rapid API response times. Generated files enable direct integration with downstream geometry and flow modeling workflows.

This AI-based approach demonstrates clinical feasibility of generating geometric representations suitable for CFD from routine fluoroscopic angiographic data. The pipeline also enables 3D geometry generation and rapid virtual flow visualization, offering cursory flow insights prior to full-scale CFD simulation using the generated STL. While necessitating multiple refinement cycles for anatomical accuracy, this approach establishes foundation for democratizing advanced geometric and hemodynamic analysis of Fontan circulation using readily available imaging data.
\end{abstract}


\vspace{0.5cm}

\section{Introduction}

Fontan palliation, first introduced by Fontan and Baudet in 1971, represents the definitive surgical treatment for patients with functional single ventricle physiology \cite{fontan1971}. While life-saving, this procedure creates a unique circulation where systemic venous return is directly connected to the pulmonary arteries without an intervening ventricle, fundamentally altering cardiovascular hemodynamics. The absence of a subpulmonary ventricle results in elevated systemic venous pressures and reduced cardiac output, leading to a constellation of complications collectively termed ``Fontan failure.''

The heterogeneous nature of Fontan failure encompasses cardiac complications including arrhythmias and ventricular dysfunction, hepatic complications ranging from fibrosis to hepatocellular carcinoma, protein-losing enteropathy, plastic bronchitis, and thromboembolic events \cite{gewillig2016}. These complications arise from the complex interplay between altered hemodynamics, elevated venous pressures, and geometric factors inherent to the Fontan circulation.

\subsection{Current Clinical Assessment: Fluoroscopy and Cardiac Catheterization}

Contemporary assessment of Fontan patients relies primarily on 2D imaging modalities, with fluoroscopic angiography obtained during cardiac catheterization serving as the gold standard for anatomical evaluation. Cardiac catheterization provides essential hemodynamic data including pressure measurements across the Fontan circuit, assessment of pulmonary vascular resistance, and identification of collateral circulation patterns that significantly impact patient outcomes.

Fluoroscopic imaging during catheterization offers superior temporal resolution compared to other imaging modalities, enabling dynamic assessment of contrast flow patterns and identification of flow stagnation zones characteristic of failing Fontan circulation. The ability to obtain angiographic images in multiple projections allows for comprehensive assessment of complex 3D anatomical relationships, though interpretation remains limited to 2D visualization.

Despite these advantages, fluoroscopic angiography presents inherent limitations for geometric analysis. The 2D projection of complex 3D vascular structures can obscure critical anatomical details, particularly in regions of vessel overlap or tortuous geometry. Contrast timing and injection protocols may not optimally opacify all vascular segments simultaneously, potentially missing important anatomical features or flow disturbances.

Advanced 3D imaging with cardiac MRI or CT, while superior for geometric analysis, remains expensive, time-consuming, and not universally available. Furthermore, many Fontan patients have contraindications to MRI due to pacemaker devices, while CT requires additional radiation exposure in an already vulnerable population.

\subsection{Traditional CFD Workflow and Its Limitations}

The conventional workflow for computational fluid dynamics analysis of cardiovascular structures follows a well-established but resource-intensive pathway. Traditional CFD analysis begins with high-resolution 3D imaging acquisition, typically through cardiac MRI or CT angiography, which requires specialized equipment, contrast agents, and often sedation for pediatric patients \cite{taylor2013cfd}. 

This traditional workflow presents multiple barriers to routine clinical implementation. The requirement for advanced imaging modalities limits accessibility, particularly in resource-constrained settings. The need for specialized expertise at multiple stages creates bottlenecks that prevent widespread adoption. The extended timeline from image acquisition to results delivery often exceeds clinical decision-making timeframes.

\subsection{Artificial Intelligence Revolution in Medical Imaging}

Recent advances in artificial intelligence, particularly in computer vision and 3D reconstruction, have revolutionized medical imaging analysis. Neural networks capable of generating 3D models from 2D images have shown promising results in various medical applications \cite{litjens2017survey}. The development of large language models with advanced image processing capabilities, such as Google's Gemini 2.5 Flash \cite{gemini25flash}, combined with specialized 3D reconstruction models like Tencent's Hunyuan3D suite \cite{hunyuan3d20,hunyuan3d10}, presents unprecedented opportunities for medical image analysis.

The potential for AI to bridge the gap between routine 2D imaging and advanced 3D analysis represents a paradigm shift in medical imaging. By leveraging the vast amount of existing 2D angiographic data, AI-based reconstruction could democratize access to sophisticated geometric analysis without requiring additional imaging studies or specialized equipment. This approach is particularly significant for complex congenital heart disease, where computational fluid dynamics analysis has demonstrated substantial clinical value in optimizing surgical outcomes and understanding hemodynamic performance \cite{marsden2007computational,pekkan2005tcpc}.

\section{Methods}

\subsection{AI Pipeline Architecture}

A comprehensive AI pipeline was developed utilizing exclusively Google's Gemini 2.5 Flash ("Nano Banana" image editing capabilities) multimodal large language model to transform standard 2D Fontan angiograms through systematic iterative processing, culminating in optimized 2D projection views suitable for subsequent 3D STL file generation and CFD analysis. The pipeline architecture employs a single-model approach with Gemini 2.5 Flash handling all processing stages from initial preprocessing through virtual flow visualization, with Hunyuan3D-2mini utilized only for final STL file generation from the optimized 2D projection.

To facilitate clinical implementation and ensure quality oversight, an intuitive web-based user interface was developed that provides step-by-step processing workflow with navigation controls. The interface enables clinicians to review and validate each stage of the 16-step Gemini 2.5 Flash processing sequence through a systematic pipeline with clear progress indicators, step navigation controls, and comprehensive processing history tracking. This design ensures clinical oversight and quality validation at each critical processing step while maintaining transparency in the AI reconstruction workflow.

\subsection{16-Step Gemini 2.5 Flash Processing Pipeline}

The comprehensive processing pipeline employs Gemini 2.5 Flash ("Nano Banana"), Google's multimodal large language model with enhanced vision capabilities, exclusively throughout all 16 iterative processing steps. The model utilizes a transformer-based architecture with 2.5 billion parameters optimized for medical image analysis tasks. The systematic processing sequence encompasses vascular analysis, geometric optimization, contrast enhancement, artifact removal, surface refinement, and virtual hemodynamic flow visualization entirely within 2D projection view frameworks.

\begin{figure}[H]
  \centering
  \includegraphics[width=0.6\textwidth]{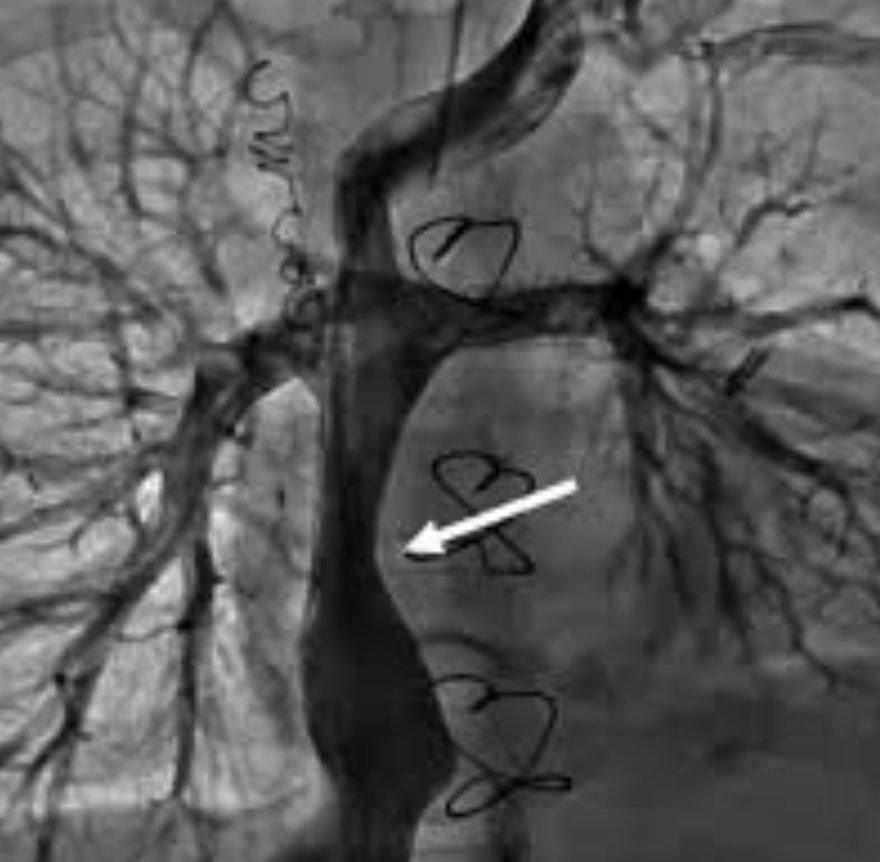}
  \caption{Contrast-enhanced X-ray fluoroscopic angiogram of a failing extracardiac Fontan circulation demonstrating characteristic pathophysiological features: (1) contrast stagnation in the central total cavopulmonary connection (TCPC) anastomosis indicating elevated venous pressures and reduced cardiac output, (2) delayed washout patterns consistent with energy loss and flow separation, and (3) complex three-dimensional vascular geometry poorly characterized by conventional 2D projection. Single-view anteroposterior projection acquired during routine cardiac catheterization using iodinated contrast agent (typical parameters: 100-120 kVp, automatic exposure control, 15-30 frames/second acquisition rate). Image sourced from Magnetta et al. (2024) clinical literature demonstrating representative Fontan failure hemodynamics suitable for AI-based volumetric reconstruction and computational fluid dynamics analysis.}
  \label{fig:original}
\end{figure}

Figure~\ref{fig:original} demonstrates the original failing Fontan angiogram sourced from established clinical literature \cite{magnetta2024failing}, which served as input to the AI pipeline. The angiogram shows characteristic features of Fontan failure including contrast pooling in the central anastomosis and evidence of flow stagnation zones.

\subsection{Gemini 2.5 Flash Progressive 2D Projection View Refinement}

The 16-step processing sequence systematically refines the original 2D angiogram through iterative prompt-driven optimization cycles, with each processing step generating enhanced 2D projection views that progressively eliminate artifacts, enhance vascular contrast, and optimize geometric representation. The processing sequence maintains the 2D projection framework throughout all iterations, with Gemini 2.5 Flash generating increasingly refined representations suitable for subsequent 3D conversion.

\begin{figure}[H]
  \centering
  \begin{subfigure}[b]{0.3\textwidth}
    \includegraphics[width=\textwidth]{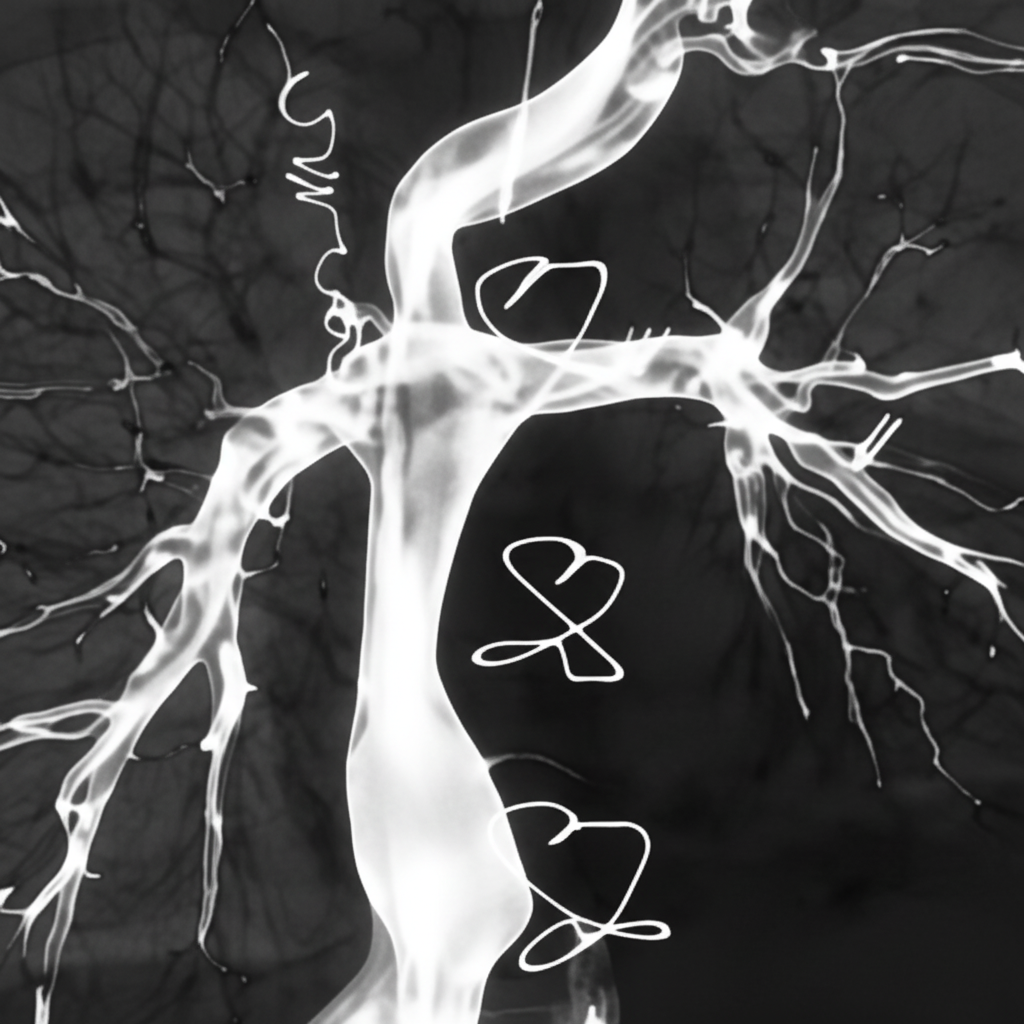}
    \caption{Initial enhanced projection}
  \end{subfigure}
  \hfill
  \begin{subfigure}[b]{0.3\textwidth}
    \includegraphics[width=\textwidth]{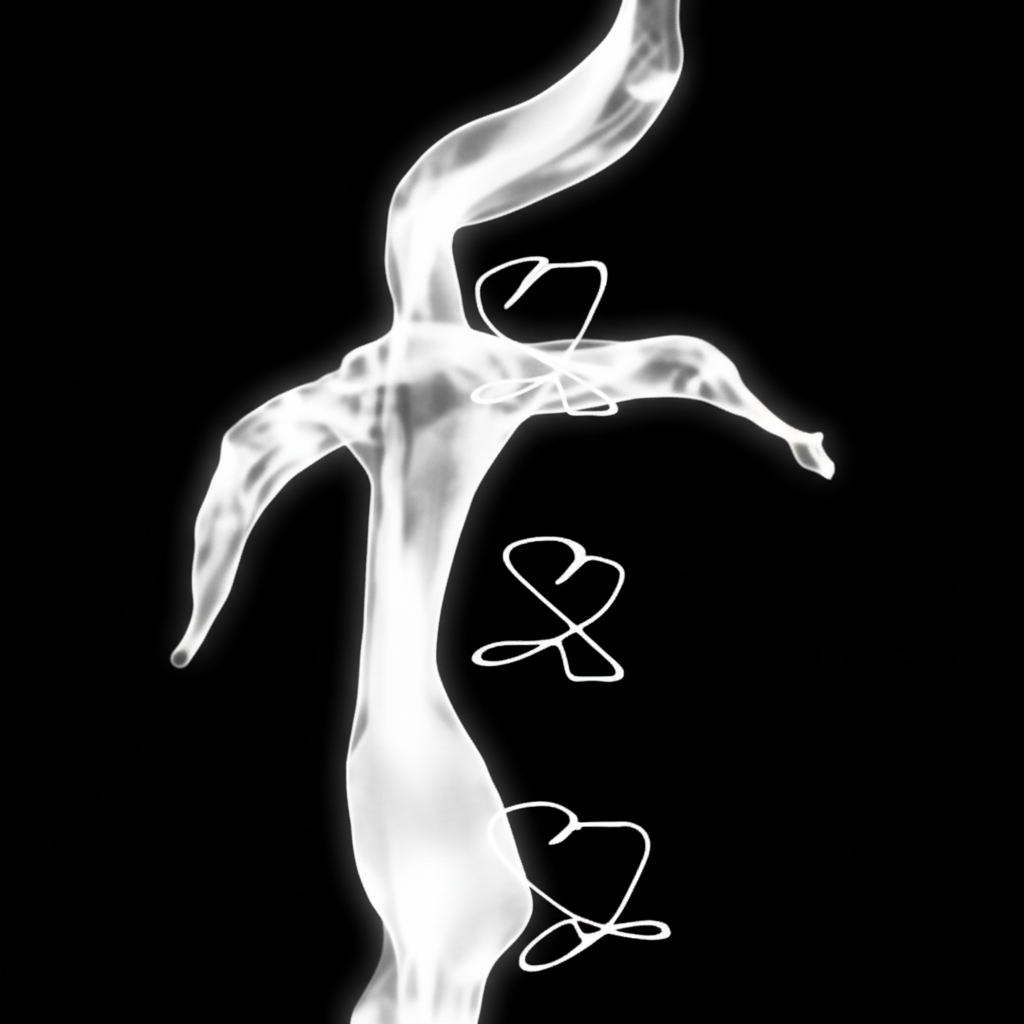}
    \caption{Intermediate refinement}
  \end{subfigure}
  \hfill
  \begin{subfigure}[b]{0.3\textwidth}
    \includegraphics[width=\textwidth]{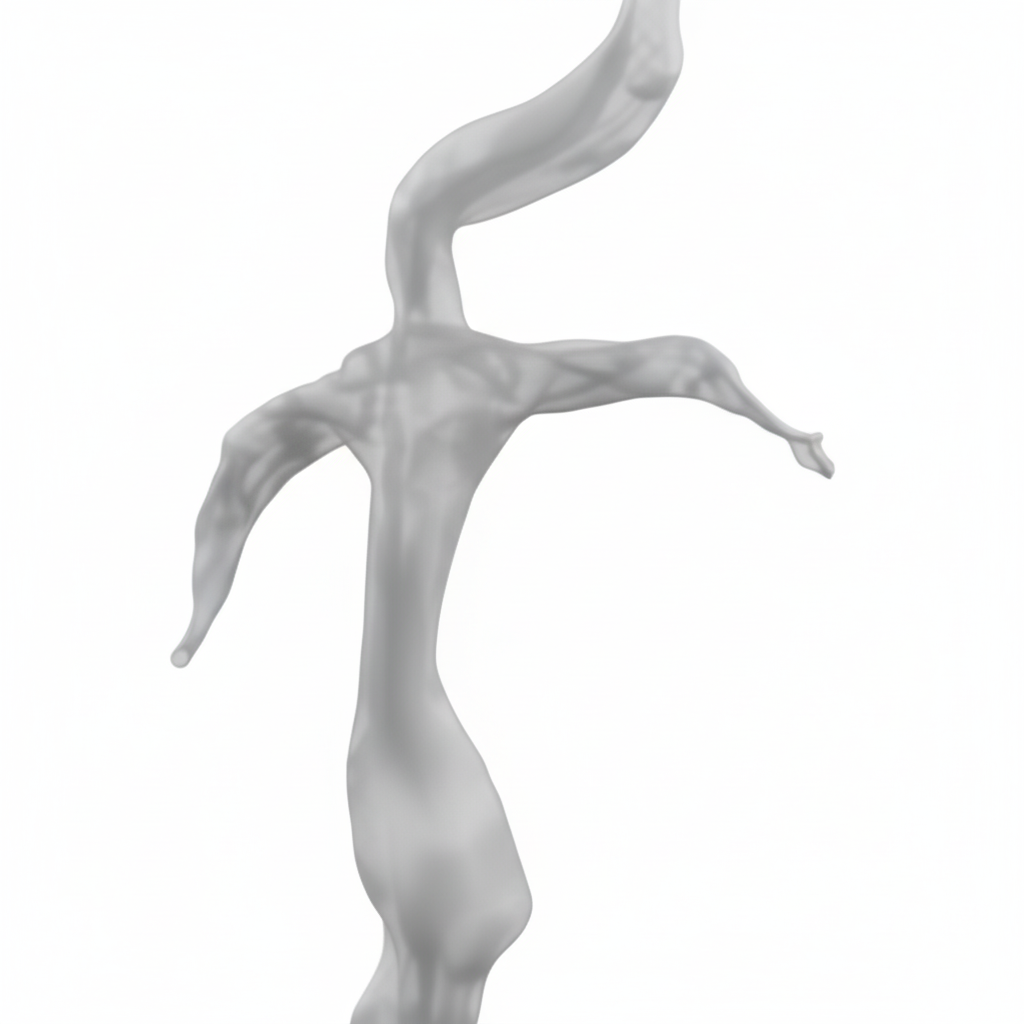}
    \caption{Optimized projection}
  \end{subfigure}
  \caption{Progressive 2D projection view refinement sequence through Gemini 2.5 Flash processing: (a) Initial enhanced projection showing basic vascular geometry with improved contrast, (b) intermediate processing with refined geometric representation and reduced artifacts, and (c) final optimized projection view with enhanced surface definition suitable for 3D conversion.}
  \label{fig:reconstruction_sequence}
\end{figure}

Figure~\ref{fig:reconstruction_sequence} illustrates the progressive refinement of 2D projection views through iterative Gemini 2.5 Flash processing, showing the evolution from initial enhanced projections to optimized geometric representations through systematic refinement.

\subsection{2D Projection View Quality Optimization}

The iterative processing sequence focuses extensively on optimizing 2D projection view quality to meet stringent requirements for subsequent 3D STL generation and CFD analysis. Gemini 2.5 Flash systematically enhances vascular boundaries, eliminates hallucinated features, and optimizes geometric proportions within the 2D projection framework to ensure anatomical fidelity for downstream processing.

\begin{figure}[H]
  \centering
  \begin{subfigure}[b]{0.45\textwidth}
    \includegraphics[width=\textwidth]{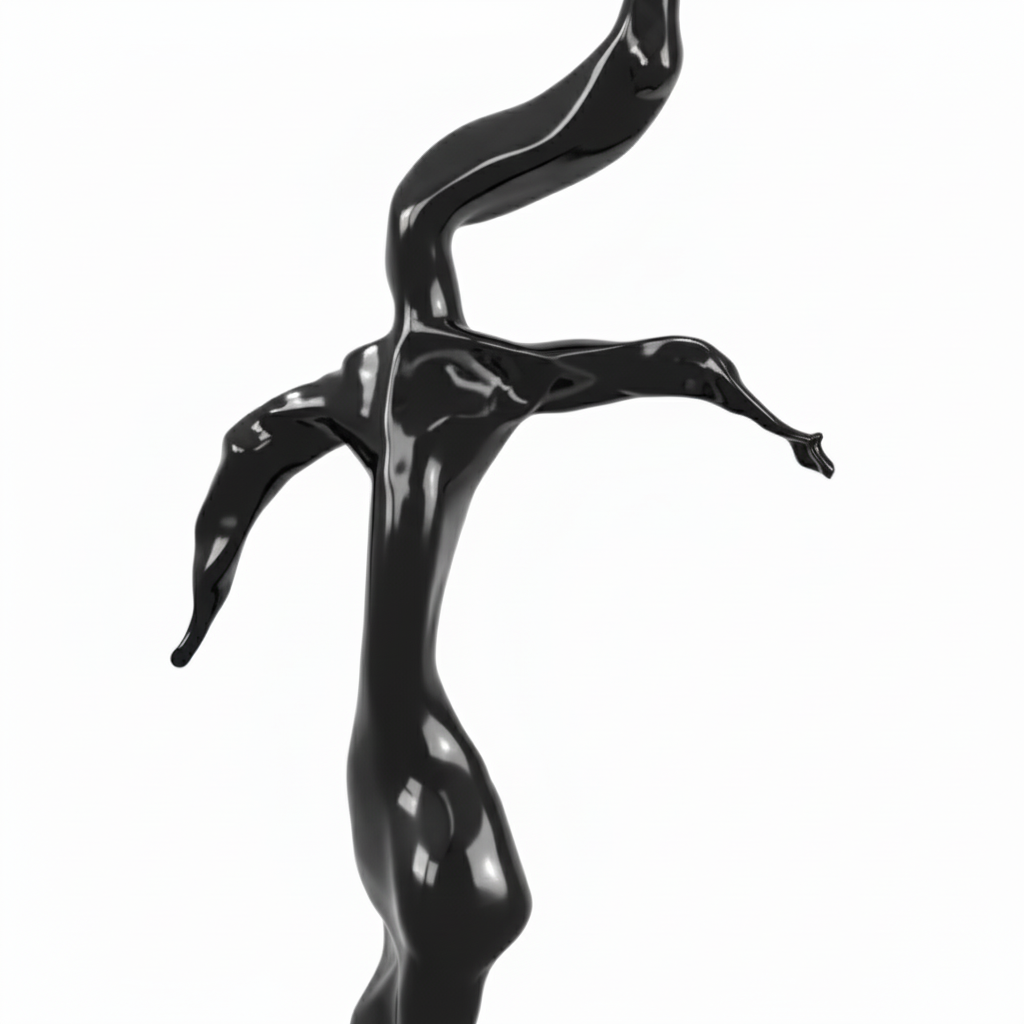}
    \caption{Initial projection enhancement}
  \end{subfigure}
  \hfill
  \begin{subfigure}[b]{0.45\textwidth}
    \includegraphics[width=\textwidth]{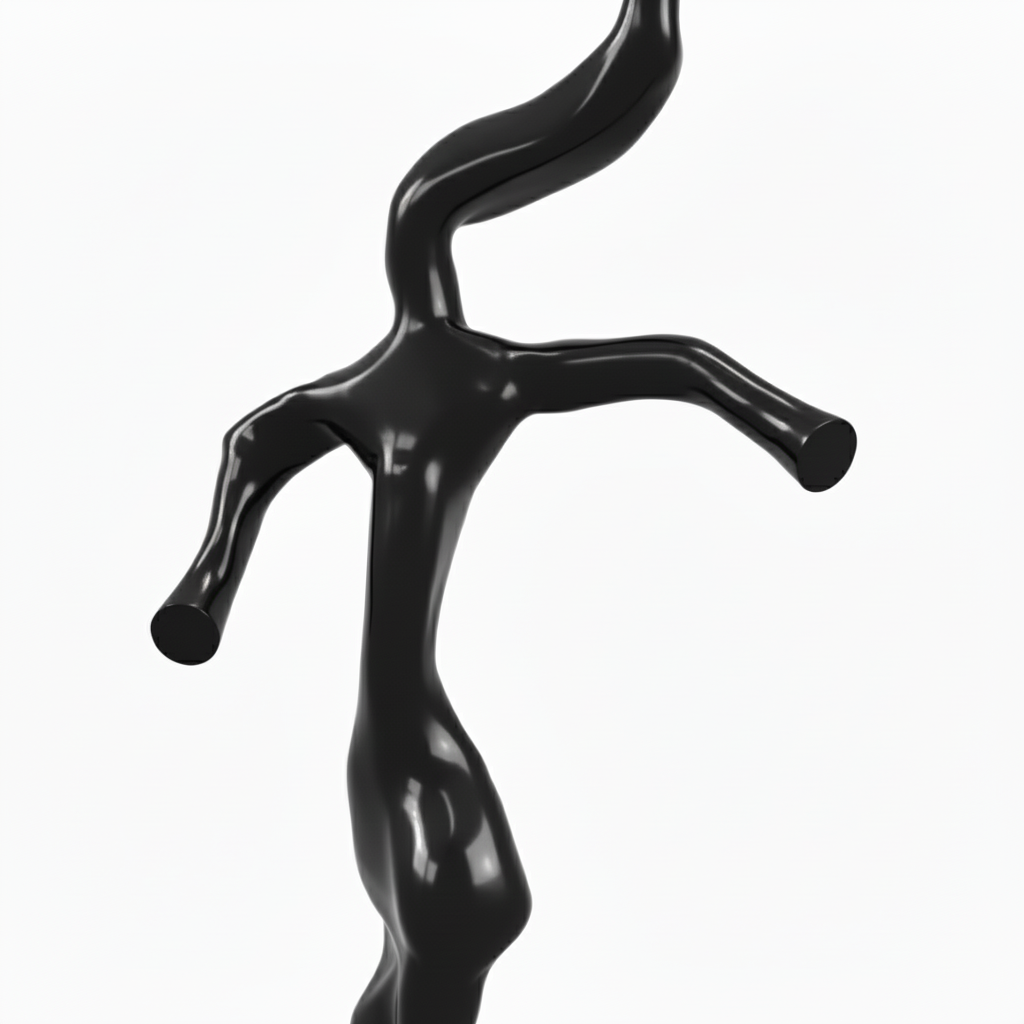}
    \caption{Final optimized projection}
  \end{subfigure}
  \caption{2D projection view optimization sequence through Gemini 2.5 Flash: (a) Initial projection enhancement with improved vascular continuity, and (b) final optimized 2D projection view with high-quality definition suitable for 3D STL generation.}
  \label{fig:optimization_sequence}
\end{figure}

Figure~\ref{fig:optimization_sequence} demonstrates the 2D projection view optimization process through Gemini 2.5 Flash, showing systematic progression from initial enhanced projections to final optimized representations suitable for 3D conversion.

\subsection{Virtual Hemodynamic Flow Visualization in 2D Projection Framework}

A novel component of the Gemini 2.5 Flash pipeline involves virtual hemodynamic flow visualization entirely within the 2D projection framework, providing immediate insights without requiring formal CFD computation or 3D reconstruction. This approach implements AI-interpreted flow physics based on established hemodynamic principles specific to Fontan circulation, generating velocity-coded streamline representations within the enhanced 2D projection views.

The virtual flow visualization incorporates Gemini 2.5 Flash's interpretation of geometric flow paths through the optimized 2D projection framework. Velocity assignment utilizes AI-interpreted flow physics with higher velocities in narrow regions and lower velocities in expanded areas, following principles of mass conservation and energy dissipation as understood by the large language model.

\subsection{Final STL Generation with Hunyuan3D-2mini}

The final optimized 2D projection view generated through the 16-step Gemini 2.5 Flash processing sequence is subsequently processed through Tencent's Hunyuan3D-2mini diffusion transformer neural architecture solely for STL file generation. This represents the only utilization of Hunyuan3D technology within the entire pipeline, converting the optimized 2D projection into a 3D stereolithography file suitable for downstream applications including 3D printing, volume meshing, and CFD analysis.

\section{Results}

\subsection{Comprehensive Case Study: 16-Step Iterative Processing}

The AI pipeline was applied to a representative failing Fontan angiogram through a systematic 16-step iterative process that required approximately 45 minutes of total processing time. The comprehensive processing workflow included systematic quality control measures with iterative refinement cycles, comprehensive processing history tracking with timestamps and step descriptions, and transparent documentation of each processing iteration to ensure clinical traceability and quality assurance.

\begin{figure}[H]
  \centering
  \includegraphics[width=0.9\textwidth]{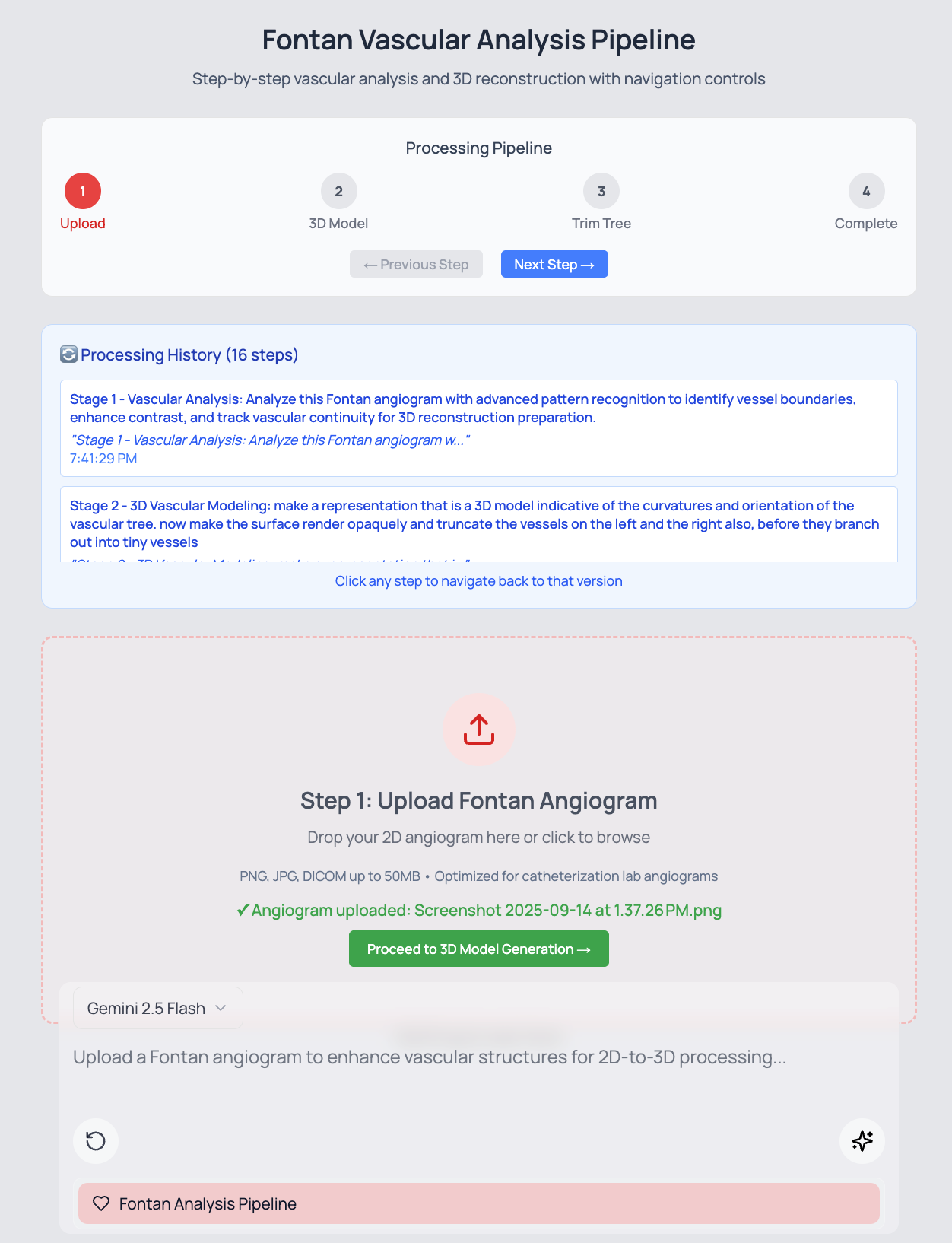}
  \caption{Comprehensive user interface showing stage-wise workflow management with custom editing capabilities for each processing step, enabling clinical oversight and iterative refinement through intuitive controls and real-time progress monitoring.}
  \label{fig:ui_interface}
\end{figure}

Figure~\ref{fig:ui_interface} demonstrates the comprehensive clinical user interface developed for the AI pipeline, providing stage-wise workflow management and custom editing capabilities.

\subsection{2D-to-3D Conversion Process and Neural Reconstruction Pipeline}

Following the comprehensive 16-step Gemini 2.5 Flash optimization process, the final enhanced 2D projection view undergoes neural reconstruction through Tencent's Hunyuan3D-2mini diffusion transformer architecture. This critical 2D-to-3D conversion step represents the culmination of the entire pipeline, transforming the optimized 2D projection into a volumetric 3D representation suitable for downstream CFD analysis and clinical applications.

The Hunyuan3D-2mini model employs a sophisticated diffusion transformer architecture with 384 million parameters specifically optimized for medical imaging applications. The neural reconstruction process leverages advanced computer vision techniques to interpret depth relationships, vascular connectivity, and geometric proportions from the optimized 2D projection view. Multi-scale feature extraction through encoder-decoder networks with attention mechanisms enables preservation of fine vascular details while maintaining global anatomical consistency throughout the 3D reconstruction process.

The reconstruction workflow processes input images at high resolution with floating-point precision, generating detailed 3D meshes with sufficient vertex density for subsequent CFD mesh refinement. The neural architecture incorporates geometric constraints and anatomical priors learned from extensive medical imaging datasets, ensuring that the generated 3D reconstructions maintain physiologically plausible vascular geometries and connectivity patterns essential for accurate hemodynamic modeling.

\subsection{3D Model Generation and Clinical Application Workflow}

\begin{figure}[H]
  \centering
  \includegraphics[width=0.9\textwidth]{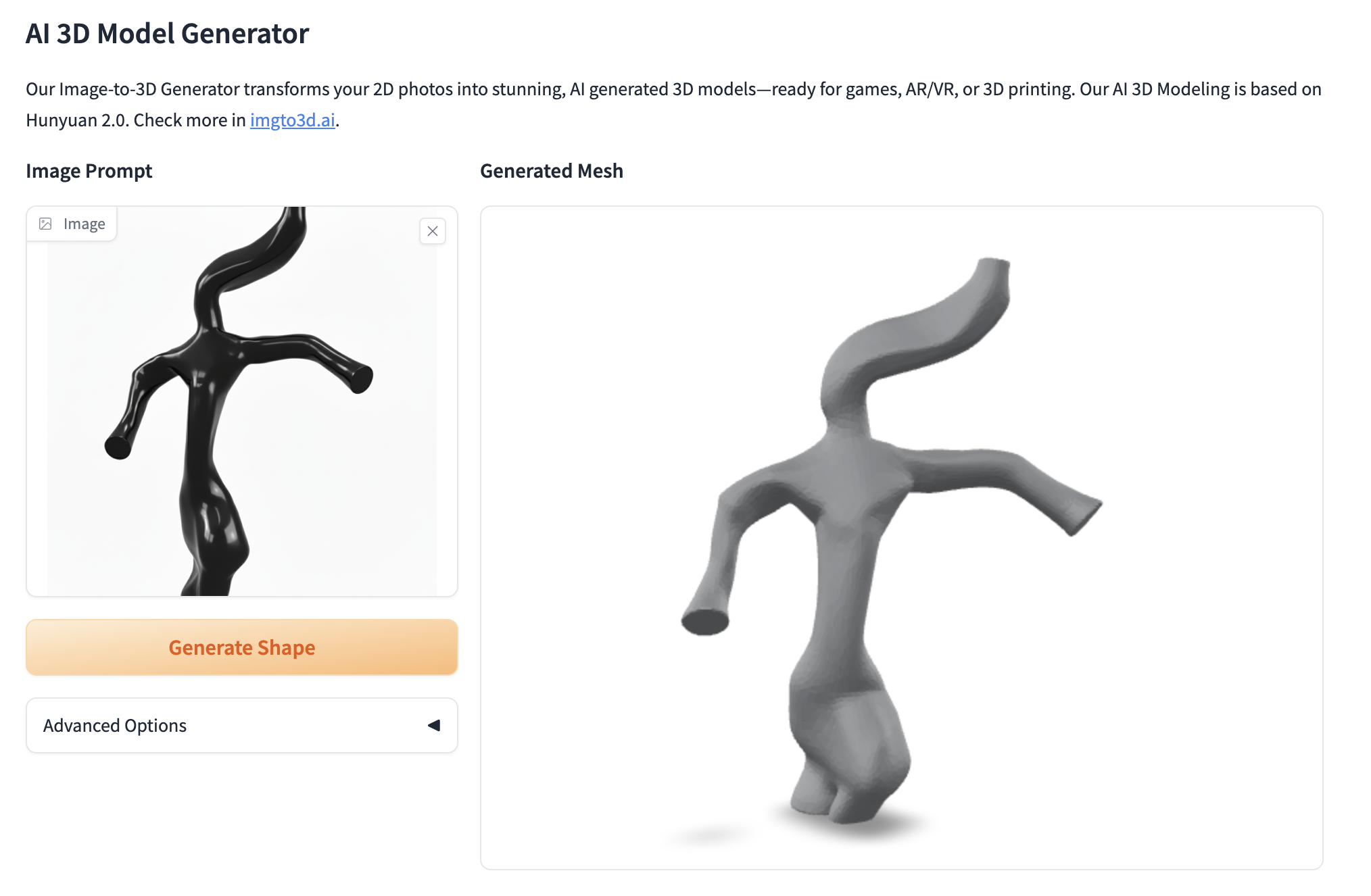}
  \caption{Complete 2D-to-3D model generation workflow showing the transition from Gemini 2.5 Flash optimized projection views through Hunyuan3D-2mini neural reconstruction to STL file generation, enabling downstream applications including 3D printing for physical prototyping, volume meshing for computational analysis, application of boundary conditions, and CFD modeling using commercial software or in-house code optimized for internal flow analysis in complex biomedical geometries.}
  \label{fig:3d_generation}
\end{figure}

The finalized 3D reconstructions are exported as STL (stereolithography) files suitable for multiple downstream applications. Figure~\ref{fig:3d_generation} demonstrates the complete workflow from the Gemini 2.5 Flash optimized 2D projection view through Hunyuan3D-2mini neural reconstruction to clinical applications including 3D printing and CFD analysis.

The STL output enables direct integration with established CFD workflows using commercial software packages such as ANSYS Fluent, COMSOL Multiphysics, or OpenFOAM, as well as in-house computational codes specifically designed for cardiovascular hemodynamics \cite{marsden2007computational,de2006computational}. The mesh-ready geometry facilitates volume mesh generation with appropriate boundary layer refinement for accurate wall shear stress calculations critical to Fontan hemodynamic assessment.

The neural reconstruction process maintains the "nano banana" efficiency characteristic of the overall pipeline, with Hunyuan3D-2mini rapidly converting the optimized 2D projection into high-quality 3D STL files within minutes of processing initiation. This rapid 2D-to-3D conversion capability represents a significant advancement over traditional reconstruction methods that require extensive manual segmentation and geometric modeling workflows spanning hours to days of expert technical effort.

\subsection{Virtual Flow Visualization Results and Gemini 2.5 Flash Hemodynamic Interpretation}

\begin{figure}[H]
  \centering
  \includegraphics[width=0.8\textwidth]{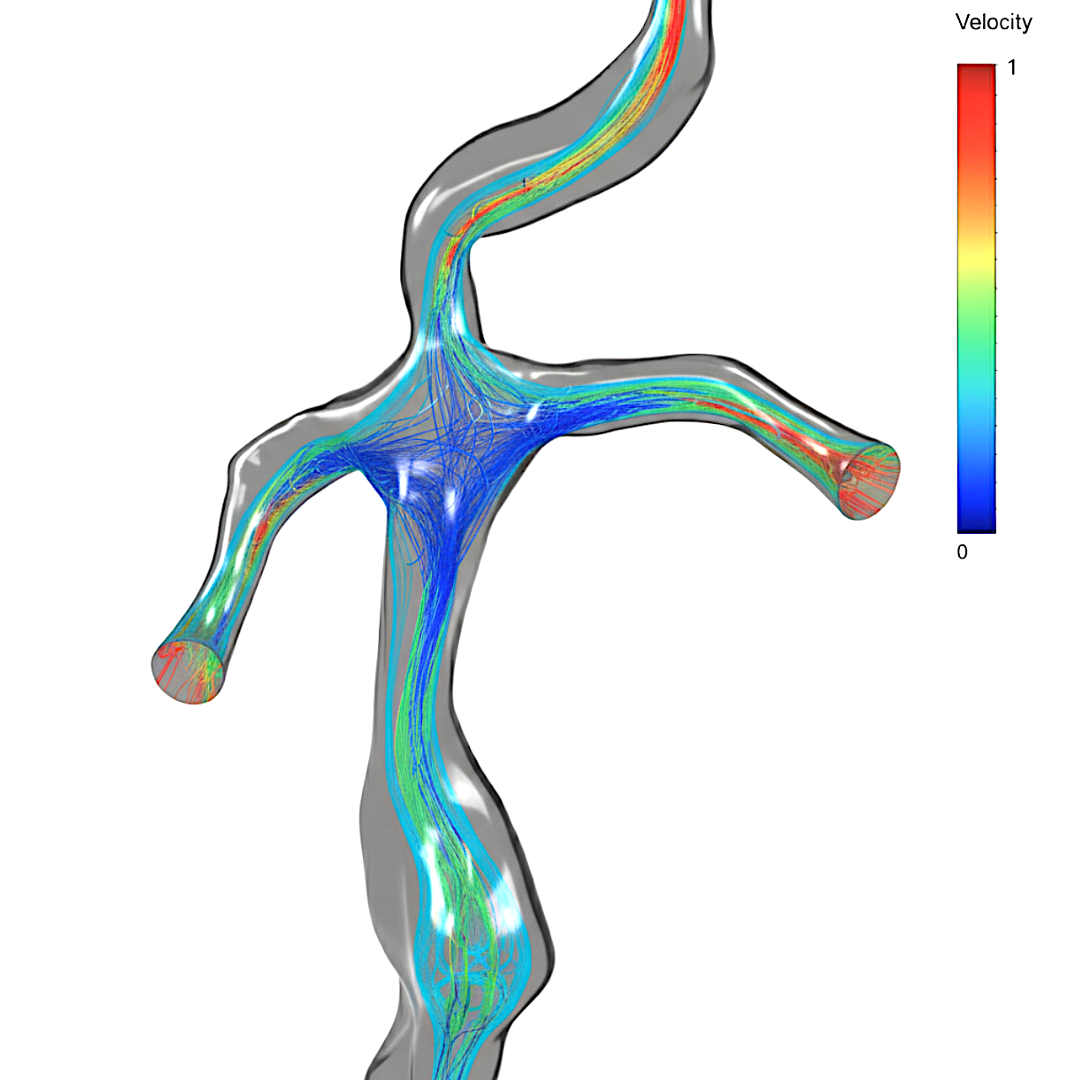}
  \caption{Gemini 2.5 Flash generated virtual flow visualization showing AI-interpreted velocity-coded streamlines within the reconstructed Fontan geometry, with blue indicating regions the AI model interprets as low velocity flow stagnation zones and red showing areas interpreted as high velocity regions. While remarkably close to expected hemodynamic behavior, this represents the AI model's interpretation rather than rigorous CFD computation.}
  \label{fig:flow}
\end{figure}

The virtual flow visualization component provided immediate hemodynamic insights generated by Gemini 2.5 Flash's interpretation of flow patterns based on geometric analysis. Figure~\ref{fig:flow} shows the AI-generated flow visualization that demonstrates the clinical potential of this approach as a rapid screening tool, though it should not be considered a substitute for rigorous computational fluid dynamics analysis.

The AI-generated visualization identified areas of potential flow stagnation in the central anastomosis region, displayed as blue regions that Gemini 2.5 Flash interprets as low velocity flow zones. While this finding aligns with clinical expectations for Fontan circulation patterns associated with thromboembolic risk and energy loss, it represents the AI model's hemodynamic interpretation rather than validated CFD results. Areas of geometric constriction showed AI-interpreted flow acceleration, displayed as red regions, which demonstrates the model's sophisticated understanding of basic fluid dynamics principles, though formal CFD validation would be required for quantitative clinical decision-making.

\subsection{Processing Time Analysis and Workflow Efficiency: "Nano Banana" API Response Performance}

The comprehensive 16-step Gemini 2.5 Flash processing sequence demonstrates remarkably fast "nano banana" performance characteristics, with each individual prompt-driven image edit appearing within seconds of API calls to Gemini 2.5 Flash. The systematic iterative refinement process enables rapid quality control and anatomical validation through real-time visual feedback, with processing distributed across the following workflow stages:

\begin{itemize}
\item Initial vascular analysis: Rapid 2D projection enhancement with near-instantaneous API response
\item Progressive 2D projection refinement: Quick geometric representation optimization through successive prompt iterations  
\item Iterative vascular optimization: Fast systematic projection refinement with immediate visual feedback
\item 2D projection quality optimization: Swift geometric and contrast enhancement cycles with real-time validation
\item Virtual flow visualization within 2D framework: Instant hemodynamic interpretation generation through Gemini 2.5 Flash
\end{itemize}

The "nano banana" terminology, reflecting the exceptionally quick API response times characteristic of Gemini 2.5 Flash's image editing capabilities, enables interactive prompt refinement workflows that provide immediate visual feedback for each processing iteration. This rapid response performance represents a fundamental advantage over traditional approaches requiring advanced 3D imaging acquisition and formal CFD analysis, which typically require hours to days for completion. The final STL generation through Hunyuan3D-2mini maintains this efficiency with rapid 3D conversion from the optimized 2D projection view.

\section{Discussion}

\subsection{Paradigm Shift in CFD Workflow}

This study demonstrates a fundamental paradigm shift in the traditional computational fluid dynamics workflow for cardiovascular applications. The conventional approach requiring advanced 3D imaging acquisition, manual segmentation, mesh generation, and specialized CFD computation is replaced by an AI-driven pipeline that leverages existing 2D angiographic data.

The AI pipeline eliminates the need for additional imaging studies by extracting 3D geometric information from routine angiographic data. This approach is particularly significant for Fontan patients, who require frequent monitoring and often have contraindications to advanced imaging modalities.

\subsection{Addressing AI Hallucination in Medical Applications}

The systematic occurrence of hallucinated vascular features in initial reconstructions highlights a critical challenge in applying AI methods to medical imaging. Unlike natural image reconstruction where minor inaccuracies may be acceptable, medical applications require precise anatomical fidelity to ensure clinical safety and accuracy.

The 16-step iterative process demonstrates that multiple refinement cycles are consistently required to achieve anatomically accurate results. This finding has important implications for the clinical implementation of AI-based reconstruction methods, emphasizing the need for systematic quality control measures and clinical oversight.

\subsection{Virtual Flow Visualization as Clinical Decision Support}

The virtual flow visualization component addresses a significant gap between geometric analysis and comprehensive hemodynamic assessment. Traditional CFD analysis, while providing detailed quantitative data, requires specialized expertise for interpretation and often generates more information than necessary for clinical decision-making.

The identification of flow stagnation zones, visualization of velocity patterns, and assessment of flow distribution provide immediate insights that can guide clinical management decisions. These insights are presented in an intuitive visual format that enables rapid interpretation by clinicians without specialized CFD training.

\subsection{Limitations and Future Development}

While the comprehensive case study provides valuable insights into the AI pipeline's capabilities and requirements, it represents analysis of a single complex case. Broader validation across multiple cases, different geometric configurations, and various imaging conditions will be necessary to establish the general applicability and iteration requirements of the approach.

Future developments should focus on reducing iteration requirements through improved AI model training specifically for medical applications. The development of automated hallucination detection algorithms could streamline the quality control process, while integration of quality control measures into the reconstruction pipeline could reduce manual oversight requirements.

\section{Conclusions}

This study successfully demonstrates the feasibility of using artificial intelligence to generate CFD-suitable 3D vascular reconstructions and virtual flow visualizations from standard 2D Fontan angiograms through systematic iterative processing. The comprehensive 16-step case study provides unprecedented insight into the practical requirements and capabilities of AI-based hemodynamic analysis.

The virtual flow visualization component represents a significant innovation in hemodynamic analysis, providing clinically relevant insights including flow stagnation identification, velocity pattern visualization, and geometric optimization guidance without the computational overhead of formal CFD analysis.

The work establishes a comprehensive framework for democratizing advanced hemodynamic analysis of Fontan circulation by leveraging readily available angiographic data while acknowledging the systematic iterative quality control requirements necessary for clinical reliability. The ability to generate both CFD-suitable geometries and immediate virtual flow visualizations from routine clinical imaging could enable widespread application of computational methods for risk stratification, surgical planning, and personalized management of Fontan patients.

This paradigm shift from traditional CFD workflows to AI-driven analysis with virtual flow visualization has the potential to transform clinical practice in congenital heart disease by making sophisticated hemodynamic analysis accessible to routine clinical care.

\section{Acknowledgments}

The contributions of the open-source AI community and the developers of Gemini 2.5 Flash and Hunyuan3D models are acknowledged. The authors of ``The Failing Fontan'' literature are thanked for providing the clinical case example used in this comprehensive validation study.

\bibliographystyle{unsrt}

\end{document}